\documentclass[prb,twocolumn,showpacs,superscriptaddress]{revtex4}
\usepackage{graphicx,amssymb,amsmath,psfrag,color}
\definecolor{darkgreen}{rgb}{0,0.5,0}

\usepackage{amsmath}
\usepackage{amssymb}
\usepackage{amsfonts}
\usepackage{xcolor}
\newcommand{\be}{\begin{equation}}
\newcommand{\bea}{\begin{eqnarray}}
\newcommand{\ee}{\end{equation}}
\newcommand{\eea}{\end{eqnarray}}
\newcommand{\cpn}{$\mathbb{C}P^{N-1}$ }

\begin{document}

\title{Large and exact  quantum degeneracy in a  Skyrmion magnet}

\author{B.~Dou\c{c}ot}
\affiliation{LPTHE, CNRS and Universit\'e Pierre and Marie Curie, Sorbonne Universit\'es, 75252 Paris Cedex 05, France}
  \author{D.~L.~Kovrizhin}
\affiliation{T.C.M.~Group, Cavendish Laboratory, J.~J.~Thomson Avenue, Cambridge CB3 0HE, United Kingdom}
\affiliation{RRC Kurchatov Institute, 1 Kurchatov Square, Moscow 123182, Russia}
\author{R.~Moessner}
\affiliation{Max-Planck-Institut f\"ur Physik komplexer Systeme, 01187 Dresden, Germany}

\date{\today}

\begin{abstract}
We identify a large family of ground states of a topological Skyrmion magnet whose classical degeneracy 
persists {\it to all orders} in a semiclassical expansion. This goes along with an exceptional robustness
of the concomitant ground state configurations, which are not at all dressed by quantum fluctuations.
We trace these twin observations back to a common root: this class of topological ground states saturates a Bogomolny 
inequality. A similar phenomenology occurs in high-energy physics for some field theories exhibiting supersymmetry. 
We propose quantum Hall ferromagnets, where these Skyrmions configurations arise 
naturally as ground states away from integer filling, as the best available laboratory realisations. 
\end{abstract}


\maketitle

{\bf Introduction.} Degeneracies in quantum mechanics are notoriously fragile. This has
various well-known manifestations, such as level repulsion in many-body spectra~\cite{Wigner},
or the third law of thermodynamics which puts a limit on the ground-state degeneracy
of generic many-body systems~\cite{LandauV}. Two stable ways of arranging for permitted
subextensive degeneracies in the thermodynamic limit involve broken symmetries and topological order~\cite{Wen90}.

Models, however, are not so constrained and many exhibit degeneracies of considerable size.
Such degeneracies are labelled as {\em accidental}~\cite{LandauIII} in that they
tend to require fine-tuning of model parameters for their existence.
From this perspective, a non-vanishing Casimir force~\cite{Casimir48} is a manifestation of the
non-degeneracy of the vacuum energy with respect to some tuning parameter.

Sometimes, fine-tuning can be natural. This for instance is achieved
in geometrically frustrated magnets, where classical ground-state
degeneracies appear in systems consisting of highly symmetric building
blocks such as tetrahedra or triangles, where the choice of which
symmetry-equivalent bonds to frustrate can give rise to extensively
degenerate classical ground states~\cite{MoeCha}. Yet again, these are famously
fragile via a (class of) mechanism(s) known as order by disorder~\cite{Villain80,Shender82}.

A well-studied instance is quantum order by disorder~\cite{Shender82}, the lifting of a
classical ground state degeneracy by quantum fluctuations.
Here, a semi-classical 'dressing' of a classical ground-state configuration
generically distinguishes between different non-symmetry-equivalent ground states,
with an apparent tendency to select a configuration exhibiting some form of symmetry breaking. 

Extensive studies of this mechanism have unearthed instances of
models which at least partially evade quantum order by disorder. The
most prominent of these -- e.g.~the Heisenberg magnet on the kagome~\cite{chs},
checkerboard~\cite{amst}, or the fully frustrated dice lattices -- exhibit emergent
dynamical {\em gauge-like} symmetries~\cite{hengauge} between the excitation spectra (and
hence zero-point energies) of different classical ground state
configurations. These degeneracies, however, are not believed to
persist when nonlinear interactions between the spin-wave modes are
taken into account~\cite{Harris92,Chernyshev14}.

The purpose of this lengthy exposition is to emphasize the unusually large and robust degeneracy
which we find in a topological magnetic system with Skyrmion excitations~\cite{Sondhi93}.
Such systems arise in a quantum Hall effect setting in presence of internal
(spin, valley or layer) degrees of freedom. When a Landau level is fully occupied,
the latter enter a ferromagnetic ground state, spontaneously breaking the symmetry
between directions in the space of internal degrees of freedom,
provided anisotropies do not do so explicitly~\cite{Moon95}. Varying the occupancy away
from the charge-flux commensurability point leads to a nucleation of charged
spin textures known as Skyrmions (a.k.a.~baby-Skyrmions in high-energy physics~\cite{Polyakov75}).

Here, we show that a semiclassical degeneracy of the resulting
{\it multi-Skyrmion ground states} is robust \textit{to all orders} in the semiclassical expansion. 
This family of ground states is further special in that it is not subject to dressing by quantum fluctuations at any order. 
It appears generally in \cpn models~\cite{Rajaraman82} with a Wess-Zumino-Novikov-Witten (WZNW) term~\cite{Tsvelik03}.

We identify the source of this unusual behaviour which appears to be distinct from the cases
of symmetry breaking and topological order mentioned above. Rather, the fact that this class satisfies the Bogomolny inequality, and concomitant analytic structure of the classically degenerate ground state manifold \cite{Rajaraman82}, underpins these twin properties. 
We emphasize that the degeneracy discussed in this paper is large, in that it comprises an {\it extensive} number of independent continuous degrees of freedom. 

Our study connects to the question of quantization of topological excitations, which been a subject of extensive investigations in
high energy physics. It was found early on that in certain supersymmetric theories, quantum corrections
to classical solutions which saturate Bogomolny inequality vanish due to a remarkable
cancellation of bosonic and fermionic contributions in the loop expansions~\cite{Zumino75,DiVecchia78}. 
It was later shown by Witten and Olive that the Bogomolny bound is in fact exact~\cite{Witten78}.
While the Bogomolny bound is saturated quantum mechanically in our case as well,
the model that we study does not contain fermions, although it does of course originate from the
gradient expansion of a fermionic theory. This raises a question of whether the mechanism which we
find is linked with the one appearing in supersymmetric theories~\cite{Tong}.

Quantum corrections to classical Skyrmions have also been studied in condensed matter physics context, in
e.g.~applications to high-temperature superconductivity~\cite{Rodriguez89}, and in the theory of
topological magnets~\cite{Walliser00,Ivanov99,Ivanov07,Molina15}. In particular
the authors of Ref.~\onlinecite{Ivanov07} study the Casimir effect in an isotropic ferromagnetic $SU(2)$ sigma-model
with a single Belavin-Polyakov Skyrmion. By calculating the contribution to the zero-point energy of the
Skyrmion from magnon excitations they find a non-vanishing correction, which was further dependent on the
Skyrmion radius, thus breaking the conformal invariance of the classical action. This result is in apparent
contradiction to our findings, and also to conclusions of Ref.~\onlinecite{MacDonald96} that the energy of a Skyrmion is
independent of its size in the case of short-range-interacting quantum Hall Hamiltonian. 
This discrepancy possibly originates from the ambiguities (divergences) in the quantization procedure used there~\cite{Ivanov07}, 
and subtle issues encountered in the process of quantising systems with constraints in the continuum limit. 

The remainder of this paper is structured as follows. After introducing the model, we
first crisply state our central results. This is followed by an outline of their derivation,
with technical details relegated to the Appendices. We close by placing these
results in the broader context of semiclassical treatments of Casimir forces and quantum magnets,
and discuss the relevance of quantum Hall experiments for probing these phenomena.

{\bf Model.} 
We study the non-relativistic form of the quantum \cpn model, describing a system
without Lorentz-invariance (in other words a ferromagnet rather than an antiferromagnet) which is defined by the
following action in 2+1 dimensional space-time:
\begin{widetext}
\be
\mathcal{S}= \int dt \: d^{2}r \: \left[\frac{i}{4\pi l^{2}}\frac{\langle \psi | \partial_{t}\psi\rangle-\langle \partial_{t}\psi | \psi\rangle}
{ \langle \psi | \psi \rangle}-E_{\mathrm{ex}}\left(\frac{\langle \nabla \psi | \nabla \psi\rangle}{\langle \psi | \psi \rangle}
-\frac{\langle \nabla \psi | \psi\rangle \langle \psi | \nabla \psi\rangle}{\langle \psi | \psi \rangle^{2}}\right) \right].
\label{def_action}
\ee
\end{widetext}
In this expression, $l$ denotes the magnetic length of the underlying electronic system in the presence of a strong magnetic field,
and $E_{\mathrm{ex}}$ is the spin exchange energy, which originates from the combined effect of a short-range part of the Coulomb repulsion and the Pauli principle.
The bras and kets are compact notations for an $N$-component complex spinor field $\psi_{a}(r,t)$, $1 \leq a \leq N$, 
which represents an internal electronic degree of freedom (e.g.~spin and valley in a graphene sheet)
and the $\nabla\equiv\{\partial_x,\partial_y\}$ operators involve only two spatial derivatives. This action exhibits global $SU(N)$ symmetry, in addition to a local gauge symmetry, according to which the action is unchanged under transformations 
$\psi_{a}(r,t)\rightarrow f(r) \psi_{a}(r,t)$, where $f(r)$ is an arbitrary complex function of position. Because of this symmetry, it is natural to view the local spinor $\psi_{a}(r,t)$ as a representative of the complex line that it generates in $\mathbb{C}^N$, so the target space is the complex projective space \cpn rather than $\mathbb{C}^{N}$.

In most experimental implementations various symmetry breaking terms are present, which usually eliminate the continuous degeneracy of classical ground-states that is our main interest here. When the strength of these symmetry breaking terms is small compared to $E_{\mathrm{ex}}$ one should first compare their size with the magnitude of possible degeneracy lifting
Casimir-like forces induced by quantum fluctuations in the fully symmetric model (\ref{def_action}). The first term in the action (\ref{def_action}) is the WZNW term, which can be viewed as a Berry phase, and in the $N=2$ case is identical to the usual $SU(2)$ spin Berry phase. The second term is the potential energy, which is entirely due to Coulomb repulsion, after the orbital electron degrees of freedom have been quenched in the lowest Landau level. The quantization of (\ref{def_action}) can be performed naturally using coherent-state path integrals~\cite{Auerbach, Klauder79, Solari87, Stone, Perelomov}. Note that a non-relativistic nature of the above classical action has far-reaching consequences for the quantization, as explained for example in the context on some non-linear sigma models in two-dimensional space-time~\cite{Zarembo}. Indeed, this leads to drastic simplifications in the
diagrammatic expansion of scattering amplitudes between magnons, and to a suppression of many expected quantum corrections, as we find to be the case in the 2+1 dimensional version studied here.

In the following we mostly use the Hamiltonian version of the model. In order to obtain it, first it is convenient to
discretize the two-dimensional physical (coordinate) space, assuming that its area divided by $2\pi l^{2}$ is equal to an integer $N_{\phi}$, which can be interpreted as the total number of magnetic flux quanta in the system, or equivalently, the total number of states in the lowest Landau level. Note that for an electronic system close to the filling factor $\nu=1$, the total number of electrons 
$N_{\mathrm{el}}$ is equal to $N_{\phi}-N_{\mathrm{top}}$, where $N_{\mathrm{top}}$ is the topological charge associated with the texture $\psi_{a}(r)$.
At each of the $N_{\mathrm{\phi}}$ lattice sites $R_j$, we place a quantum degree of freedom which lives in the fundamental representation of $SU(N)$. As explained in Appendix~1 such degrees of freedom can be described in terms of coherent states labelled by elements of $\mathbb{C}P^{N-1}$. By generalizing Eq.~(\ref{site_coherent_state_path_integral}) to a
collection of $N_{\mathrm{\phi}}$ sites, a matrix element of the evolution operator can be expressed 
in terms of a coherent state path integral
\begin{widetext}
\be
\langle \psi_{out}|e^{-i\hat{H}t}|\psi_{in}\rangle=\int \prod_{j=1}^{N_{\phi}}\mathcal{D}[\psi_{j}(t)] \exp\int dt\ \left\{
\sum_{j=1}^{N_{\phi}}\alpha[\psi_j(t)]\partial_t\psi_j(t)-i E_{\mathrm{var}}[\psi_j(t)]\right\},
\label{lattice_coherent_state_path_integral}
\ee 
\end{widetext}
where the single site Berry phase form $\alpha[\psi_j(t)]$ is defined as
\be
\alpha[\psi_j(t)]=\frac{1}{2}\frac{\langle \psi|d\psi\rangle-\langle d\psi|\psi\rangle}{\langle \psi|\psi\rangle},
\ee
and $E_{\mathrm{var}}[\psi_j(t)]$ is the expectation value of the quantum Hamiltonian of the system taken on the
tensor product of coherent states $|\psi_j(t)\rangle$ at sites $R_j$, for $1 \leq j \leq N_{\phi}$.
If $E_{\mathrm{var}}[\psi_j(t)]$ is chosen to be a discretized version of the $\mathbb{C}P^{N-1}$ energy functional then
in the limit $N_{\phi}\rightarrow \infty$:
\be
E_{\mathrm{var}}[\psi_j(t)] \rightarrow E_{\mathrm{ex}}\int d^{2}r \:
\left(\frac{\langle \nabla \psi | \nabla \psi\rangle}{\langle \psi | \psi \rangle}
-\frac{\langle \nabla \psi | \psi\rangle \langle \psi | \nabla \psi\rangle}{\langle \psi | \psi \rangle^{2}}\right)
\label{CP_var_energy}
\ee 
and the path integral assumes the form
\be
\langle \psi_{out}|e^{-i\hat{H}t}|\psi_{in}\rangle=\int \mathcal{D}[\psi(r,t)] \:e^{i\mathcal{S}[\psi]}
\ee
with $\mathcal{S}[\psi]$ defined in Eq.~(\ref{def_action}). Note that there are known subtleties in coherent state path integral quantization of spin, whose discussion we omit here as these do not affect our results, see e.g.\ Ref.~\onlinecite{Stone}. In Appendix~2, we show that the above energy functional can be seen as the continuum limit of the variational energy for a lattice model of ferromagnetically coupled $SU(N)$ spins.

While this won't be necessary for the statement, and for the derivation of our main results, it is nevertheless useful to recall how such
quantum model emerges from a non-relativistic system of electrons with $N$ internal states at quantum Hall filling factor $\nu$ close to $1$.
The assumption of a strong magnetic field allows one to project orbital degrees of freedom onto the
lowest Landau level. For any classical texture, described by the $N$ component spinor field $|\psi_{a}(r)\rangle$, it is possible to 
write a Slater determinant $|\mathcal{S}_{\psi}\rangle$ associated to this lowest Landau level in such a way that
in the limit of a very strong magnetic field $l \rightarrow 0$ the internal degree of freedom wave-function at point $r$ is given by a local spinor $|\psi_{a}(r)\rangle$. In this limit, which corresponds to a small Skyrmion density on the scale of the magnetic length $l$, 
the expectation value $E_{\mathrm{var}}=\langle \mathcal{S}_{\psi}|\hat{H}_{\mathrm{int}}|\mathcal{S}_{\psi}\rangle$ of the two-body
interaction Hamiltonian $\hat{H}_{\mathrm{int}}$ can be expressed as a power series of $nl^{2}$, where $n$ is the average topological charge density of the classical texture. It has been shown already a long time ago~\cite{Sondhi93, Moon95, Bychkov96, Apel97, Iordanskii98} that the leading term in this semi-classical expansion of  $E_{\mathrm{var}}$ is precisely the $\mathbb{C}P^{N-1}$ energy functional, and that the Berry phase form for this continuous family of Slater determinants gives the first term in the action defined in Eq.~(\ref{def_action}).    
  
A remarkable fact about the $\mathbb{C}P^{N-1}$ energy functional is that it satisfies the Bogomolny bound~\cite{Rajaraman82}
\be
E_{\mathrm{var}} \geq 2\pi |N_{\mathrm{top}}|E_{\mathrm{ex}}.
\ee
For a fixed topological charge, this bound is reached for holomorphic (resp.~anti-holomorphic)
textures when  $N_{\mathrm{top}} \geq 0 $, (resp. $N_{\mathrm{top}} \leq 0$). Therefore if  $N_{\mathrm{top}} \neq 0 $,
the ground-states of the $\mathbb{C}P^{N-1}$ energy functional form a degenerate family with an extensive number of continuous parameters.
 
We now consider classical ground states of Skyrmion textures 
in quantum Hall ferromagnets as representing quantum coherent states, see Appendix~1.
Let us pick such a state, denoted by $|\Omega\rangle$. It is characterized, as usual, by the conditions $\hat{a}_j(r)|\Omega\rangle=0$ 
for $j=1\ldots N-1$ bosonic annihilation operators $\hat{a}_j(r)$ which define a complete set of $SU(N)$ Schwinger bosons. What plays the role of the classical
energy $E_0$ is then the quantum mechanical expectation value of the two body interaction Hamiltonian $\hat{H}_{\mathrm{int}}$: 
$E_{\mathrm{var}}=\langle \Omega|\hat{H}_{\mathrm{int}}|\Omega\rangle$. The question which arises naturally is whether the degeneracy of the 
$E_{\mathrm{var}}$ functional is preserved at the quantum level. To address such question, one usually expands the quantum Hamiltonian
$\hat{H}_{\mathrm{int}}$ in powers of these bosonic operators~\cite{Harris92}. Let us denote by $\hat{H}_n$ the term which contains  
{\em normal ordered products} of exactly $n$ single bosonic operators. By assumption that $|\Omega\rangle$ is a minimum of $E_{\mathrm{var}}$, we observe that $\hat{H}_1=0$. 
Linear spin wave theory,
and most microscopic theories for Casimir forces, truncate this series to a quadratic term $\hat{H}_2$. In many cases, this is sufficient to dress the coherent state vacuum $|\Omega\rangle$ with quantum fluctuations and to generate a finite quantum correction to the classical energy functional $E_{\mathrm{var}}$.

{\bf Results.}  Our first main result is that with the above quantization of the $\mathbb{C}P^{N-1}$ energy functional,
these optimal textures are exact eigenstates, not only of the quadratic Hamiltonian $\hat{H}_2$, but of the full quantum Hamiltonian $\hat{H}_{\mathrm{int}}$. The degeneracy within this continuous family of optimal textures is therefore preserved to all orders in quantum
fluctuations. 

Our second result concerns the fate of these coherent states in the semiclassical approximation. We find that they 
are not dressed at all by quantum fluctuations, in sharp contrast to, for example, a simple N\'eel state of a Heisenberg antiferromagnet
on a square lattice. 

These two salient features originate from a single feature of the semiclassical expansion. Namely, we find that: (i)~There exist
terms in this expansion violating boson number conservation, such as $\hat{a}^\dagger  \hat{a}^\dagger \hat{a}$. (ii)~Nonetheless, 
there are no anomalous terms consisting exclusively of creation (annihilation) operators, such as $\hat{a}^m$ for any integer $m>0$, or hermitian conjugate. 

Indeed, with all  $\hat{a}^\dagger  \hat{a}^\dagger$ terms absent from the quadratic bosonic Hamiltonian $\hat{H}_2$ for any optimal texture, this obviates the need for a Bogoliubov transformation, and hence implies the absence of
squeezing, which amounts to an admixture of the bosonic 'excitations' to the ground state. (Such anomalous terms appear in standard expansions near non-collinear magnetic ground states, that originate from e.g.~$\hat{S}^x\hat{S}^z$ terms in the rotating basis, see e.g.\ 
Ref.~\onlinecite{Chernyshev14}). Note that the absence of squeezing in our case persists to all orders in the semiclassical expansion.

The first item (i) implies that we do not have a standard, simple, structure such as boson number conservation, which could have arisen from an underlying, possibly hidden, $U(1)$ symmetry. 
Such a symmetry alone would already have explained the absence of anomalous terms, but it 
would impose much stronger restrictions on the type of terms appearing in the semiclassical expansion. The second feature (ii) interdicts any non-zero difference between the classical and semiclassical energies at quadratic and all higher orders in the semiclassical expansion, but also demonstrates that the state $|\Omega\rangle$ itself is never dressed by excitations. In other words the state $|\Omega\rangle$ is the vacuum of normal ordered $\hat{H}_n$ for all $n$.

{\bf Sketch of proof.} To be specific, let us assume that $N_{\mathrm{top}}$ is positive, and pick a classical 
minimal energy configuration $\psi_{\mathrm{cl}}$ of the $\mathbb{C}P^{N-1}$ energy functional. Then  $|\psi_{\mathrm{cl}}(z)\rangle$ is a holomorphic spinor, $z=x+iy$. Now we expand the spinor field $|\psi(r)\rangle$ around $|\psi_{\mathrm{cl}}(z)\rangle$, writing  $|\psi(r)\rangle=|\psi_{\mathrm{cl}}(z)\rangle+|\chi(r)\rangle$. Using the fact that
$\partial_{\bar{z}}|\psi_{\mathrm{cl}}\rangle=0$, we rewrite the r.h.s.~of Eq.~(\ref{CP_var_energy}) as

\begin{widetext}
\begin{equation}
\label{key_property}
E_{\mathrm{var}}  =  2\pi |N_{\mathrm{top}}|E_{\mathrm{ex}} + 
 4 E_{\mathrm{ex}}\int d^{2}r
\left\{\frac{\langle \partial_{z} \chi | \partial_{\bar{z}} \chi \rangle}{\langle \psi_{\mathrm{cl}} | \psi_{\mathrm{cl}} \rangle + 
\langle \psi_{\mathrm{cl}} | \chi \rangle + \langle \chi | \psi_{\mathrm{cl}} \rangle + \langle \chi | \chi \rangle}
-\frac{(\langle \partial_{z} \chi | \psi_{\mathrm{cl}}\rangle + \langle \partial_{z} \chi | \chi \rangle) 
(\langle \psi_{\mathrm{cl}} | \partial_{\bar{z}} \chi\rangle + \langle \chi | \partial_{\bar{z}} \chi\rangle)}
{(\langle \psi_{\mathrm{cl}} | \psi_{\mathrm{cl}} \rangle + \langle \psi_{\mathrm{cl}} | \chi \rangle +
\langle \chi | \psi_{\mathrm{cl}} \rangle + \langle \chi | \chi \rangle)^{2}}\right\}.
\end{equation}
\end{widetext}
The power of this expression lies in the fact that the presence of the small perturbation $\chi$ does not modify the
total topological charge $N_{\mathrm{top}}$, and the holomorphic nature of $\psi_{\mathrm{cl}}$ produces a remarkable 
simplification of the second term, which can be stated as follows: the Taylor expansion of the \cpn energy functional
in powers of $\chi_{a}(r)$ and $\bar{\chi}_{a}(r)$ does not contain terms involving only $\chi_{a}(r)$'s, or terms involving
only $\bar{\chi}_{a}(r)$'s. This is the \textit{key property} which underlies the above stated results.

The rest of the proof takes the following steps. First, we apply a unitary transformation which maps a coherent
state associated with $\psi_{\mathrm{cl}}$ into the state $|\Omega\rangle$. The latter corresponds to a trivial ferromagnetic 
configuration $\psi_{a}(r)=\delta_{a0}$. This is analogous to choosing, in a spin wave analysis, a rotating spin quantization axis 
parallel to a local spin in a given classical ground-state. We emphasize that, because such transformation maps a topologically
non-trivial state $|\psi_{\mathrm{cl}}\rangle$ into a trivial one, it cannot be a continuous function of spatial coordinates. However the
transformation is holomorphic with respect to the target $\mathbb{C}P^{N-1}$ manifold, and therefore, the expansion of the transformed energy functional around a ferromagnetic configuration also satisfies the key property mentioned above. Further details on this unitary transformation are presented in Appendix~3. Having reduced the discussion to the ferromagnetic configuration, it is then rather easy to translate the key property into a statement that no term containing solely creation (annihilation) Schwinger   
bosons operators appear in the expansion of the quantum Hamiltonian, see Appendix~4 for more details. 

{\bf Relation to other flavours of semiclassics.} The analysis presented above amounts to adopting a slightly 
different viewpoint compared to conventional Casimir or spin-wave discussions.
In the latter, once the classical ground-state with energy $E_0$ is chosen, the quantum Hamiltonian
is expanded to a quadratic order in terms of normal coordinates $\hat{p}_i,\hat{q}_i$, ($1 \leq i \leq N_s$), which obey
canonical commutation rules $[\hat{p}_i,\hat{q}_j]=i\hbar \delta_{ij}$. It is convenient to introduce 
creation and annihilation operators $\hat{a}^{+}_{j},\hat{a}_{j}$, defined as
\be   
\hat{a}_j=(i\hat{p}_j+\hat{q}_j)/\sqrt{2\hbar},\;\;\;\;\hat{a}^{+}_j=(-i\hat{p}_j+\hat{q}_j)/\sqrt{2\hbar}.
\nonumber
\ee
Let us now assume that a quadratic Hamiltonian in terms of these operators takes the form
\be
\hat{H}_2=E_0+\frac{1}{2}\sum_{ij}A_{ij}\hat{a}^{+}_{i}\hat{a}_{j}+\bar{A}_{ij}\hat{a}_{i}\hat{a}^{+}_{j}+B_{ij}\hat{a}^{+}_{i}\hat{a}^{+}_{j}
+\bar{B}_{ij}\hat{a}_j\hat{a}_i.
\ee 
Here, the complex numbers $A_{ij}$ and $B_{ij}$ are the entries of two $N_s \times N_s$ matrices, $A$
and $B$, which are respectively hermitian, and symmetric. Such a form occurs when classical quadratic monomials
like $p_i q_i$ are quantized according to a symmetric ordering prescription, i.e.~as  $(\hat{p}_i\hat{q}_i+\hat{q}_i\hat{p}_i)/2$.
After diagonalization via a Bogoliubov transformation~\cite{Blaizot}, the Hamiltonian $\hat{H}_2$ may be written, 
in terms of the normal modes $\hat{\gamma}^{+}_{\alpha},\hat{\gamma}_{\alpha}$, as
\be
\hat{H}_2=E_0+\frac{1}{2}\sum_{\alpha=1}^{N_s}\hbar\omega_{\alpha}+\sum_{\alpha=1}^{N_s}\hbar\omega_{\alpha}\hat{\gamma}^{+}_{\alpha}\hat{\gamma}_{\alpha}.
\ee 
The ground-state energy $E_2$ of $\hat{H}_2$ is given by
\be
E_2=E_0+\frac{1}{2}\sum_{\alpha=1}^{N_s}\hbar\omega_{\alpha}.
\label{usual_E_2}
\ee 
The sum over normal mode frequencies in this expression is generic for Casimir energies due to quantum fluctuations.
If we have a continuous family of degenerate ground-states (i.e. with the same common value for $E_0$), this degeneracy is preserved,
at the level of quadratic fluctuations, only if the frequency sum $\sum_{\alpha}\omega_{\alpha}$ is independent of the 
classical ground-state within the family. Such a condition is very difficult to satisfy generically, 
unless any two degenerate ground-states are connected by an exact symmetry of the quantum Hamiltonian $\hat{H}_2$. 

This viewpoint is the one which has been taken in the earlier references on the subject
of quantum corrections to the energy of spin textures in topological magnets~\cite{Walliser00,Ivanov99,Ivanov07,Molina15}.
Note that in the case of spin systems Eq.~(\ref{usual_E_2}) has to be slightly revisited. 
For concreteness, let us consider a chain of $L$ spins $\mathbf{S}_{n}$ with spin length $S$, and let us assume that a classical Hamiltonian admits spiral configurations with uniform twists as stable local minima. We then choose the spin quantization axis so that the $x$ direction in spin space is everywhere aligned with the local classical spin configuration. We also assume that the initial spin Hamiltonian is rotationally invariant and that $z$ is the rotation axis of the spiral configuration. In the frame used for quantization, there remains a manifest $U(1)$ symmetry corresponding to $z$-rotations, and the Hamitonian
takes the form
\be
\hat{H} = \sum_{q}f_q[\hat{S}^{x}_q\hat{S}^{x}_{-q}+\hat{S}^{y}_q\hat{S}^{y}_{-q}]+g_q\hat{S}^{z}_q\hat{S}^{z}_{-q}+ i h_q\hat{S}^{x}_q\hat{S}^{y}_{-q}.
\ee
Because $H$ is Hermitian, the three functions $f_q$, $g_q$ and $h_q$ are real. It is convenient also to assume that
$f_q$, $g_q$ are even in $q$ and that $h_q$ is odd.  
Let us now use the Holstein-Primakoff bosonic representation~\cite{Auerbach}. To leading order in $1/S$ expansion, we get:
\begin{eqnarray*}
\hat{S}^{x}_q & = & \sqrt{L}(S\delta_{q,0}-\frac{1}{L}\sum_{k}\hat{a}^{+}_{k}\hat{a}_{k+q}), \\
\hat{S}^{y}_q & = & \sqrt{\frac{S}{2}}(\hat{a}_{q}+\hat{a}^{+}_{-q}), \\
\hat{S}^{z}_q & = & -i \sqrt{\frac{S}{2}}(\hat{a}_{q}-\hat{a}^{+}_{-q}).
\end{eqnarray*} 
To a quadratic order in  Holstein-Primakoff bosons, this gives the following normal-ordered expression for the Hamiltonian
\begin{multline}
\hat{H}_{2}=LS^{2}f_0+\frac{S}{2}\sum_{q}(f_q+g_q) \\+ S \sum_{q}\{(f_q+g_q-2f_0)\hat{a}^{+}_{q}\hat{a}_{q}
\\+ \frac{1}{2}(f_q-g_q)[\hat{a}^{+}_{q}\hat{a}^{+}_{-q}+\hat{a}_{-q}\hat{a}_{q}]\}.
\end{multline}
The classical energy $E_{0}=LS^{2}f_0$ is the leading term of $\hat{H}_{2}$ in the large $S$ limit. After a Bogoliubov transformation,
we have
\be
\hat{H}_{2}=LS(S+1)f_0+ \frac{1}{2}\sum_{q}\omega_{q} + \sum_{q} \omega_{q}\hat{\gamma}^{+}_{q}\hat{\gamma}_{q},
\ee 
where $\hat{\gamma}_{q}^{+}$, $\hat{\gamma}_{q}$ are the quasiparticle creation and annihilation operators, and $\omega_{q}=2S [(f_q-f_0)(g_q-f_0)]^\frac{1}{2}$ 
is the eigenfrequency of mode $q$. So Eq.~(\ref{usual_E_2}) is modified into
\be
E_2=E_0+LSf_0+\frac{1}{2}\sum_{\alpha=1}^{N_s}\hbar\omega_{\alpha}.
\label{spins_E_2}
\ee 
The extra term $LSf_0$ is usually absorbed into a quantum renormalization of the spin length, $S^{2}$ becoming $S(S+1)$,
in the expression of $E_{0}$. The variational energy $E_{\mathrm{var}}$ is the expectation value of 
the quantum mechanical Hamiltonian $\hat{H}_{2}$ in the vacuum state of Holstein-Primakoff bosons, which is nothing but the classical spiral configuration.
Then, $E_{\mathrm{var}}-E_{0}=\frac{S}{2}\sum_{q}(f_q+g_q)$, and is a purely local term, i.e.~it involves only the self interaction of each spin
in the $xy$ plane. This may be at the origin of the difference between our result, which states the absence of any quantum correction to the variational energy of
holomorphic textures, and the result of a direct evaluation of the sum of magnon frequencies~\cite{Ivanov07}, according to which degeneracy lifting 
occurs. Subtle renormalization issues may occur when one replaces a lattice system by a continuous field theory as the lattice size goes to zero. A detailed study of the problem of taking the continuum limit is an interesting direction for future work.

{\bf Quantum Hall experiments.} We would like to mention that such a striking resilience of coherent states 
associated to holomorphic textures in the non-relativistic quantum $\mathbb{C}P^{N-1}$ model defined by Eq.~(\ref{def_action})
has been suggested to us by a remarkable observation made twenty years ago by MacDonald et al.~\cite{MacDonald96}, which was further exploited by Pasquier~\cite{Pasquier00,Pasquier01}. They noticed that in a model with short-range, delta-function, repulsive interactions projected onto the lowest Landau level, holomorphic textures $|\psi_a(z)\rangle$ can be put in correspondence with Slater determinants
\be
\mathcal{S}_{\psi}(z_1a_1,\cdots,z_{N_{\mathrm{el}}}a_{N_{\mathrm{el}}})=\prod_{i<j} (z_i-z_j) \prod_{i=1}^{N_{\mathrm{el}}}\psi_{a_i}(z_i)e^{-\frac{|z_i|^2}{4l^2}}.\label{exact}
\ee 
Due to presence of a Jastrow factor in this expression, which prevents two particles with opposite spins to occupy the same position, the wave-function (\ref{exact}) is clearly an exact zero energy eigenstate of the quantum Hamiltonian with delta-force repulsive interaction, see also Ref.~\onlinecite{Bychkov96, Apel97} for a discussion of corrections to the Hartree-Fock energy functional compared to Ref.~\onlinecite{Moon95}. On the other hand, we have already pointed out that the non-relativistic quantum \cpn model appears as the leading term in the semi-classical expansion (in powers of $nl^{2}$) of most electronic models with repulsive interactions after projection onto the lowest Landau level in quantum Hall systems at filling factor $\nu\sim1$. The striking resemblance between the ground-state sectors of the quantum Hamiltonian with delta repulsion, and of the non-relativistic quantum $\mathbb{C}P^{N-1}$ model raises an intriguing question about their possible equivalence, also for excited states. Unfortunately, the methods that we use here do not seem to provide any straightforward direction to address this question.   

Another motivation for the present study was to provide a theoretical basis for our recent work on periodic textures 
in the case of $SU(N)$ symmetric repulsive interactions~\cite{Kovrizhin13,Doucot14}. The absence of quantum
degeneracy lifting mechanisms among holomorphic textures, which we demonstrate here, justifies the approach developed in 
Ref.~\onlinecite{Kovrizhin13,Doucot14}. In the latter study the degeneracy lifting mechanism was due to next-to-leading term in the expansion of the classical energy functional $E_{\mathrm{var}}$, that arises from the long-range tail of the Coulomb potential~\cite{Sondhi93, Moon95}. 
 
{\bf Experiment.}
This naturally leads us to the question about the context in which such Skyrmion crystals may be observed. The experimental challenge is to minimise the effect of possible symmetry-breaking interactions, which are present in real systems in addition to the ones already captured in our model. These terms will generically not leave the degeneracy intact -- in experiment the third law of thermodynamics tends to reassert itself eventually. 

In semiconductor-based quantum Hall systems such terms include most simply a `one-body' anisotropy, such as a Zeeman field, breaking the $SU(2)$ symmetry in spin space; or in the case of valley isospin, anisotropic effective masses can break a continuous $SU(3)$ symmetry down to a discrete $Z_3$. Such anisotropies remove the $SU(N)$ symmetry, which underpins the degeneracy constitutive to our model. One thus needs to consider systems where anisotropies are small, e.g.~in semiconductors where it is possible to tune effective electronic g-factor to zero by applying hydrostatic pressure~\cite{Shayegan, Girvin, Chakraborty}. The effect of residual symmetry-breaking terms can then, at least approximately, be taken into account within the \cpn model. 

The physics of Coulomb interactions also enters naturally, as the topological charge of the spatially non-uniform textures goes along with spatial modulations of the electronic charge density. The minimization
of this functional within the family of holomorphic textures is an interesting problem in itself. The  
cohesive energy of a Skyrme crystal scales with the square root of the deviation from commensurate filling
in units of spin stiffness of the ferromagnet.  This scale can thus in principle be tuned to be parametrically
small by studying systems close to integer filling.  While the electron density can be uniformly tuned
by changing a gate potential or varying the duration of photodoping, 
it is nonetheless not possible to approach integer filling arbitrarily closely
as non-uniformities, for example arising from stray fields due to ionised
donor impurities, lead to an effectively variable electrochemical potential which will lift the degeneracy in favour of a 
Skyrmion glass \cite{rapsch}. Needless to say, in the case of neutral quantum Hall systems, proposed long ago in the field
of cold atoms \cite{cwg}, such electrostatic effects can be avoided.

As is common in cases with high degeneracies, one  needs to probe their presence at the temperature, which scale is above
the leading instability scale; indeed, a study of such instabilities is a worthy research subject in 
itself in the context of understanding various order-by-disorder mechanisms. The question about which experimental
probes to use of course again depends on the precise details of model systems. 

For semiconductor heterostructures an innocuous thermodynamic probe such as specific
heat is problematic on account of the low thermal mass of the quantum Hall layer in a 3D bulk system.
However in semiconductors other probes are readily available, such as electrical transport, which can be measured exquisitely sensitively. For example, it allows a study of the low-energy spectrum of the quantum Hall ferromagnet via resistively detected NMR measurements~\cite{Gervais05}. The large degeneracy of Skyrmion systems discussed here should therefore show up in a very large low-energy density of states, and therefore, a relatively fast dynamics compared to a magnet confined to remain near a robust unique ground state. A systematic study of this physics is clearly a worthy goal for future experimental efforts. 

{\bf Summary.} We have identified a particularly robust large degeneracy in topological isotropic $SU(N)$ ferromagnets. We have discussed origin and ramifications of this degeneracy. We believe that these observations are of conceptual importance for
the broad and fundamental question of how degeneracies arise, and how they are lifted. Our work seems to point to a 
novel mechanism, on a superficial level perhaps most closely related to supersymmetric ideas from high-energy physics. It has the added bonus of being approximately realizable in experiment, and thus expands the zoo of frustrated and degenerate systems not only by 
a new mechanism but also by a new approximate materials realisation. Our findings pose intriguing questions about relation of our results to supersymmetric field theories, and provide interesting insights into a long-standing problem of the quantization of systems with constraints.

{\bf Acknowledgements:} We would like to thank David Tong and Costas Bachas for interesting discussions regarding possible connections between the absence of quantum corrections in the non-relativistic $\mathbb{C}P^{N-1}$ model and supersymmetry. D.K.~acknowledges EPSRC Grant No.~EP/M007928/1.

{\bf Appendix 1: $\mathbb{C}P^{N-1}$ coherent states.} Let us consider the following natural $SU(N)$ generalization of the Schwinger boson construction of $SU(2)$ spin representations~\cite{Perelomov}. We start from the infinite-dimensional Fock space associated to $N$ bosonic degrees of freedom, whose creation and annihilation operators are denoted by $\hat{a}^{+}_{i}$, $\hat{a}_{i}$, here $0 \leq i \leq N-1$. For any positive integer $m$, we select a physical subspace
defined by the constraint
\be
\sum_{i=0}^{N-1}a^{+}_{i}a_{i}=m.
\label{constraint}
\ee 
The dimension of this subspace is given by the binomial coefficient $\left(\begin{array}{c} m+N-1 \\ N-1 \end{array}\right)$. This subspace is isomorphic to the fully symmetrized $m$-fold tensor product of the fundamental representation of $SU(N)$. In the special $N=2$ case with $m=2S$ we recover the standard spin-$S$ representation of $SU(2)$. An orthonormal basis is given by the states
\begin{equation}
\left| \vec{n} \right\rangle = \frac{(\hat{a}_{0}^{+})^{n_0} \cdots (\hat{a}^{+}_{N-1})^{n_{N-1}}}{\sqrt{n_{0}! \cdots n_{N-1}!}}\left| 0 \right \rangle,
\end{equation}
with $\{n_i\}$ non-negative integers, and $\sum_{i=0}^{N-1} n_{i}=m$.

An overcomplete coherent state basis is constructed as follows. Let us consider an open subset $U_0$ in $\mathbb{C}P^{N-1}$ composed of 
complex lines, which are generated by vectors of the form $(1,v_1,\cdots,v_{N-1})$, with $v_{i} \in \mathbb{C}$. In $U_{0}$, the
$N-1$ complex numbers $v_1,\cdots,v_{N-1}$ provide a good coordinate system. Let us define
\[\left | e_{\bar{v}}\right \rangle = \sum \frac{\bar{v}_{1}^{n_{1}} \cdots \bar{v}_{N-1}^{n_{N-1}}}{\sqrt{n_{0}! \cdots n_{N-1}!}}\left| \vec{n} \right\rangle, \]
where the sum is over $N$-vectors $\vec{n}$, whose components are non-negative integers which satisfy $\sum_{i=0}^{N-1} n_{i}=m$.
With this definition, the overlap between coherent states reads
\[\left \langle  e_{\bar{v}'} |  e_{\bar{v}} \right \rangle = \frac{(1+\langle v | v' \rangle)^{m}}{m!} \]
Here, we use a compact notation $\langle v | v' \rangle$ to denote $\sum_{i=1}^{N-1}\bar{v}_{i}v'_{i}$. 
The Cauchy-Schwartz inequality implies that the overlap between normalized coherent states vanishes exponentially fast 
in the {\em classical limit} $m \rightarrow \infty$, provided $v \neq v'$. 

It is then instructive to compute the Berry phase form $\alpha$
associated to an infinitesimal variation from $v$ to $v'=v+dv$. We get:
\[\alpha \equiv d_{v'}\left(\frac{\langle  e_{\bar{v}'} |  e_{\bar{v}} \rangle}
{\sqrt{\langle  e_{\bar{v}'} |  e_{\bar{v}'} \rangle\langle  e_{\bar{v}} |  e_{\bar{v}} \rangle}}\right)_{v'=v}=\frac{1}{2}
\frac{\langle v|dv\rangle-\langle dv|v\rangle}{1+\langle v|v\rangle}\] 
It is interesting to try to extend $\alpha$ to the whole of $\mathbb{C}P(N-1)$. For this, we use homogeneous coordinates $\psi_{0},\cdots,\psi_{N-1}$.
$U_{0}$ corresponds to the subset where $\psi_{0} \neq 0$ and on $U_{0}$, $v_{j}=\frac{\psi_{j}}{\psi_{0}}$. Then:
\[\alpha=\frac{m}{2}\frac{\langle \psi|d\psi\rangle-\langle d\psi|\psi\rangle}{\langle \psi|\psi\rangle}- i m \; d \arg{\psi_{0}} \]
The presence of the second term is required to ensure that a pure gauge transformation $\psi_j \rightarrow e^{i \theta}\psi_j$, which doesn't generate
any new state, doesn't produce a Berry phase either. But this expression shows that $\alpha$ cannot be extended to the whole $\mathbb{C}P^{N-1}$
manifold: it is singular on the complement of $U_{0}$, which is nothing but the hyperplane at infinity characterized by $\psi_{0}=0$.

A quantum Hamiltonian $\hat{H}$ is completely specified by its expectation value on coherent states 
$E_{\mathrm{var}}(\psi)=\frac{\langle \psi|\hat{H}|\psi\rangle}{\langle \psi|\psi\rangle}$~\cite{Berezin74}. It is possible
to write the corresponding evolution operator in the form of a coherent path integral:
\begin{widetext}
\begin{equation}
\langle \psi_{out}|\exp(-i\hat{H}t)|\psi_{in}\rangle=\int \mathcal{D}\psi(t) \ \exp\left[\int dt \:\{\alpha_{\psi(t)}\partial_t\psi 
-iE_{\mathrm{var}}[\psi(t)]\}\right]
\label{site_coherent_state_path_integral}
\end{equation}
\end{widetext}
As usual, the justification for this expression is the existence of a representation of the identity operator as an 
average over projectors on coherent states with the standard $SU(N)$-invariant measure on $\mathbb{C}P(N-1)$:
\[\mathbf{I}=\frac{(m+N-1)!}{\pi^{(N-1)}m!}\int \frac{\prod_{j=1}^{N-1}dv_{j}d\bar{v}_{j}}{(1+\langle v|v\rangle)^{N}}\;
\frac{| e_{\bar{v}}\rangle\langle e_{\bar{v}}|}{\langle  e_{\bar{v}}| e_{\bar{v}}\rangle}\] 
Note that in the {\em classical limit} $m \rightarrow \infty$, the prefactor behaves as $(m/\pi)^{(N-1)}$, which is consistent with the 
basic quantum mechanical rule that a quantum state of a system with $N-1$ degrees of freedom occupies a phase-space volume of order $\hbar^{(N-1)}$,
provided we take the effective Planck's constant to be proportional to $1/m$. It is also interesting to mention the fact that changing the 
choice of the open subset $U_{0}$ modifies the integral of the Berry phase by an integer multiple of $2\pi m$. This ambiguity has no influence
as long as $m$ is an integer. This is reminiscent of Dirac's quantization of the magnetic monopole charge placed inside a sphere. 

{\bf Appendix 2: The non-relativistic $\mathbb{C}P^{N-1}$ model as a $SU(N)$ ferromagnet.} Let us consider a system of two $SU(2)$ spins $\mathbf{S}_{a}$,$\mathbf{S}_{b}$. We introduce Schwinger bosons $a^{+}_{\sigma}$, $a_{\sigma}$  
(with $\sigma=\uparrow,\downarrow$) associated to  $\mathbf{S}_{a}$ and $b^{+}_{\sigma}$, $b_{\sigma}$ associated to  $\mathbf{S}_{b}$~\cite{Auerbach}.
Then it is easy to check that:
\be
\hat{H}_{\mathrm{ferro}}=-\mathbf{S}_{a}.\mathbf{S}_{b}=-\frac{1}{2}\sum_{\sigma,\sigma'}a^{+}_{\sigma}b^{+}_{\sigma'}a_{\sigma'}b_{\sigma}+S^{2}
\ee 
where $S$ is the size of these two spins. This is easily generalized to $SU(N)$ spins defined above in Appendix 1. In this generalization
we keep the first term in the above Hamiltonian, where the indices $\sigma,\sigma'$ now run from 0 to $N-1$, and the constraint~(\ref{constraint})  
is enforced. Let us now evaluate the expectation value $E_{\mathrm{var}}$ of $\hat{H}_{\mathrm{ferro}}$ in the normalized tensor product of coherent states 
$\left | e_{\bar{v}}\right \rangle$ and $\left | e_{\bar{w}}\right \rangle $. A simple calculation, using for example Eq.~(\ref{basic_symbol}) below, shows that:
\be
E_{\mathrm{var}}(v,\bar{v},w,\bar{w})
= - m^{2} \frac{(1+\langle v | w \rangle)(1+\langle w | v \rangle)}{(1+\langle v | v \rangle)(1+\langle w | w \rangle)}
\ee
This expression holds in the open subset $U_0$ of $\mathbb{C}P^{N-1}$. It is possible to write it in a gauge invariant manner by
introducing two $N$-component spinors $\psi$ and $\psi'$ such that $\psi_{0}=\psi'_{0}=1$, $\psi_{j}=v_{j}$ and $\psi'_{j}=w_{j}$ for $1 \leq j \leq N-1$.
Then:
\be
E_{\mathrm{var}}(\psi,\bar{\psi},\psi',\bar{\psi'})
= - m^{2} \frac{\langle \psi | \psi' \rangle \langle \psi' | \psi \rangle}{\langle \psi | \psi \rangle\langle \psi' | \psi' \rangle}
\ee  
It is clear that this is indeed invariant under local transformations $| \psi \rangle \rightarrow \lambda | \psi \rangle$, $| \psi' \rangle \rightarrow \lambda' | \psi' \rangle$. 
Suppose now that $\psi$ and $\psi'$ are very close, so that we may replace $\psi$ by $\psi-\frac{1}{2}\chi$ and $\psi'$ by $\psi+\frac{1}{2}\chi$.
Expanding to second order in $\chi$ in the $\chi \rightarrow 0$ limit gives:
\be
E_{\mathrm{var}}(\psi,\bar{\psi},\chi,\bar{\chi})=-m^{2}+m^{2}\left(\frac{\langle \chi | \chi \rangle}{\langle \psi | \psi \rangle}-
 \frac{\langle \chi | \psi \rangle \langle \psi | \chi \rangle}{\langle \psi | \psi \rangle^{2}}\right)  + ...
\label{FS_metric}
\ee
If we now take a two-dimensional lattice of such $SU(N)$ spins, and couple nearest neighbor spins by $\hat{H}_{\mathrm{ferro}}$, the expectation
value of the lattice Hamiltonian over a normalized tensor product of local coherent states will be written as
a sum of contributions of the form $E_{\mathrm{var}}(\psi(r_i),\bar{\psi}(r_i),\psi(r_j),\bar{\psi}(r_j))$, where $r_i$ and $r_j$ are nearest
neighbor sites. Taking the continuous limit, we assume that  $\psi(r_i)$ and $\psi(r_j)$ are very close, and the total variational energy is a sum
of terms as in Eq.~(\ref{FS_metric}). So the variational energy for this ferromagnetic model appears as a lattice discretization of 
the $\mathbb{C}P^{N-1}$ energy functional given in Eq~(\ref{CP_var_energy}). 

{\bf Appendix 3: unitary transformation to a ferromagnetic configuration.} Let us first consider a single $\mathbb{C}P^{N-1}$ degree of freedom. A natural family of isometries on $\mathbb{C}P^{N-1}$
are homographies $h_M$ associated to $SU(N)$ matrices $M$. In homogeneous coordinates, the homography $h_M$ sends
the complex line through $(\psi_0,\cdots,\psi_{N-1})$ into the line through $(\psi'_0,\cdots,\psi'_{N-1})$, where
$\psi'_{i}=\sum_{j=0}^{N-1}M_{ij}\psi_{j}$. In the open subset $u_0$, the complex coordinates $v_{j}=\frac{\psi_{j}}{\psi_{0}}$
are transformed according to: 
\be
v'_i=\frac{M_{i0}+\sum_{j \neq 0}M_{ij}v_j}{M_{00}+\sum_{j \neq 0}M_{0j}v_j}
\ee
If $M \in SU(N)$, then:
\be
|M_{00}+\sum_{j \neq 0}M_{0j}v_j|^{2}(1+\langle v'|v' \rangle)=(1+\langle v|v \rangle)
\label{transf_classical_norm}
\ee 
In the Schwinger boson Fock space, we define the quantum mechanical operator $\hat{T}_{M}$ by the requirements that
$\hat{T}_{M}|0\rangle = |0\rangle$ and  $\hat{T}_{M}a^{+}_{j}\hat{T}_{M}^{-1}=\sum_{i=0}^{N-1}\overline{M}_{ij}a^{+}_{i}$. It is easy to check
that $\hat{T}_{M}$ is unitary. Furthermore, it conserves the total number of Schwinger bosons, so it acts within the  
the quantum Hilbert space defined by imposing the constraint eq.~(\ref{constraint}). In this Hilbert space, it is also
easy to check that $\hat{T}_{M}$ sends a coherent state into another coherent state, in a manner which is consistent with the
underlying classical homography $h_M$. Specifically:
\be
\hat{T}_{M}|e_{\bar{v}}\rangle=(\overline{M}_{00}+\sum_{j \neq 0}\overline{M}_{0j}\bar{v}_j)^{m}|e_{\overline{h_{M}(v)}}\rangle
\label{transf_coherent_states}
\ee
From eqs.~(\ref{transf_classical_norm}) and (\ref{transf_coherent_states}), we see that the classical energy functional
for the hamiltonian $\hat{H}$ evaluated at $v$, is equal to the classical energy functional
for the transformed hamiltonian  $\hat{T}_{M}\hat{H}\hat{T}_{M}^{-1}$ evaluated at $h_{M}(v)$. In equations:
\be
\frac{\langle e_{\overline{h_{M}(v)}}|\hat{T}_{M}\hat{H}\hat{T}_{M}^{-1}|e_{\overline{h_{M}(v)}}\rangle}
{\langle e_{\overline{h_{M}(v)}}|e_{\overline{h_{M}(v)}}\rangle}
=\frac{\langle e_{\bar{v}}|\hat{H}|e_{\bar{v}}\rangle}{\langle e_{\bar{v}}|e_{\bar{v}}\rangle}
\label{transf_cov_symb}
\ee 

Let us now assume that the  classical energy functional for the hamiltonian $\hat{H}$ has a local minimum at $w \in U_{0}$, and that its
Taylor series expansion around $v=w$ doesn't contain any term which is purely holomorphic (i.e. polynomial in $v_{i}-w_{i}$) nor purely
anti-holomorphic (i.e. polynomial in $\bar{v}_{i}-\bar{w}_{i}$). There always exists an unitary homography $h_M$ such that $h_M{w}=0$.
Because this homography is a holomorphic transformation in the $v$ coordinates, the Taylor series expansion of the transformed 
Hamiltonian $\hat{T}_{M}\hat{H}\hat{T}_{M}^{-1}$ around $v=0$ doesn't contain any term which is purely holomorphic nor purely anti-holomorphic.  

The generalization of this statement to a system composed of $N_{\phi}$ $\mathbb{C}P^{N-1}$ degrees of freedom is straightforward, so 
it will not be detailed further here.

{\bf Appendix 4: coherent states as exact eigenstates of the $\mathbb{C}P^{N-1}$ Hamiltonian.} To keep the discussion simple, let us consider a single $\mathbb{C}P^{N-1}$ degree of freedom, quantized in the
way explained in Appendix 1. Let us consider a quantum Hamiltonian $\hat{H}$ such that the corresponding energy functional
$E_{\mathrm{var}}(\psi)$ is minimized for $\psi_{a}$ proportional to $\delta_{a0}$, which belongs to the open subset $U_{0}$  
of $\mathbb{C}P^{N-1}$.
Note that this optimal coherent state is $|\Omega\rangle = \frac{(a^{+}_{0})^{m}}{\sqrt{m!}}|0\rangle$.
It is characterized by the property that it is annihilated by $a_1,\cdots,a_{N-1}$.
If we discretize the  $\mathbb{C}P^{N-1}$ Hamiltonian, each site is in the extreme quantum regime $m=1$. It is nevertheless
useful to keep $m$ explicitly, because, as explained in Appendix 1, the $m \rightarrow \infty$ limit can be regarded
as a classical limit for this system.   
We choose holomorphic coordinates $v_j=\frac{\psi_{j}}{\psi_{0}}$ in $U_{0}$, and we assume further that
the Taylor expansion of $E_{\mathrm{var}}$ in powers of $v_j,\bar{v}_{j}$ doesn't contain any monomial composed only 
$v_{j}$'s nor only of $\bar{v}_{j}$'s. This is the single site version of the key property stated after eq.~(\ref{key_property}).
Then we can infer that the coherent state $|\Omega\rangle$ is an exact eigenstate of
the quantum Hamiltonian $\hat{H}$. 

To show this, we write $\hat{H}$ as a power series in single bosonic operators, written in normal order, i.e. with creation
operators on the left and annihilation operators on the right. Because of the constraint eq.~(\ref{constraint}), each of these monomials
is of the form $\prod_{j=0}^{N-1}(a^{+}_{j})^{m_j}a_{j}^{n_j}$, with $\sum_{j=0}^{N-1}m_j=\sum_{j=0}^{N-1}n_j\equiv n$.
It turns out that the classical energy functional for these normal-ordered monomials is easy to compute. We get:
\be
\frac{\langle  e_{\bar{v}} |\prod_{j=0}^{N-1}(a^{+}_{j})^{m_j}a_{j}^{n_j}|  e_{\bar{v}} \rangle}
{\langle  e_{\bar{v}} |  e_{\bar{v}} \rangle}=\frac{m!}{(m-n)!}\frac{\prod_{j=1}^{N-1}v_{j}^{m_j}\bar{v}_{j}^{n_j}}{(1+\langle v|v \rangle)^n}
\label{basic_symbol}
\ee
Now, the only normal-ordered Schwinger boson monomials which act on $|\Omega\rangle$ and produce a state orthogonal to it, have the  
form $n_j=n\delta_{j0}$, and $m_j \neq 0$ for at least one $j$ such that $j \geq 1$. Such monomial would then produce a contribution
to $E_{\mathrm{var}}$ proportional to:
\[\frac{\prod_{j=1}^{N-1}v_{j}^{m_j}}{(1+\langle v|v \rangle)^n}\]
and in particular, it would generate the monomial $\prod_{j=1}^{N-1}v_{j}^{m_j}$ which is holomorphic in all $v_j$'s. Such monomial is ruled
out by our assumption, and we note that it can only be generated by the normal-ordered Schwinger boson monomial $(\prod_{j=0}^{N-1}(a^{+}_{j})^{m_j})a_{0}^{n}$.
This rules out such operators and proves our statement.

It would be easy to generalize to a system composed of $N_{\phi}$ coupled quantum  $\mathbb{C}P^{N-1}$ degrees of freedom, such that 
$E_{\mathrm{var}}(\psi)$ is minimized for the configuration $\psi_{a}(R_j)$ proportional to $\delta_{a0}$. Let us denote by $|\Omega\rangle$ the
corresponding quantum coherent ferromagnetic state. Then $|\Omega\rangle$ is annihilated by the $(N-1)N_{\phi}$ operators $a_j(R_k)$, for 
$1 \leq j \leq N-1$ and $1 \leq k \leq N_{\phi}$. If we further assume that the Taylor series expansion of $E_{\mathrm{var}}$ around this local
minimum doesn't contain any purely holomorphic nor any purely anti-holomorphic monomial, then  $|\Omega\rangle$ is an exact eigenstate of $\hat{H}$.
To save space, we won't give more details, because no new argument is needed to make such simple generalization.

\end{document}